\documentstyle[12pt]{article}
\begin{document}
\setlength{\baselineskip}{0.30in}

\newcommand{\beq}{\begin{equation}}
\newcommand{\eeq}{\end{equation}}

\newcommand{\bi}{\bibitem}

{\hbox to\hsize{September 1995 \hfill FTUV/95-42 \hfill IFIC/95-44}}

\begin{center}
\vglue .06in
{\Large \bf { Invisible Matter }}\\
Lectures presented at XXIII ITEP Winter School, \\
Zvenigorod, March, 1995 \\
(to be published in Surveys of High Energy Physics) \\[.5in]
{\bf A.D. Dolgov \footnote{Permanent address: ITEP, 117259, Moscow,
Russia.} }
\\[.05in]
{\it{ Instituto de Fisica Corpuscular - C.S.I.C. \\
Departament de Fisica Teorica, Universitat de Valencia \\
46100 Burjassot, Valencia, SPAIN}}\\[.15in]

{Abstract}\\[-.1in]
\end{center}
\begin{quotation}
These lectures have been given to particle physicists, mostly
experimentalists and very briefly and at a pedestrian level review the
problems of dark matter. The content of the lectures is the following:
1. Introduction.
2. Cosmological background.
3. Luminous matter.
4. Primordial nucleosynthesis and the total amount of baryons.
5. Gravitating invisible matter.
6. Baryonic crisis.
7. Inflationary omega.
8. Intermediate summary.
9. Possible forms of dark matter.
10. Structure formation: basic assumptions.
11. Structure formations: basics of the theory.
12. Evolution of perturbations with different forms of dark matter.
13. Conclusion.
The presentation and conclusion reflects personal view of the author
that a considerable amount of invisible energy in the universe is in the
form of vacuum energy (cosmological constant) and possibly in the form
of a classical field which adjusts vacuum energy to the value
permitted and requested by astronomical data.
\end{quotation}

\newpage
\section{Introduction}

There are very strong indications that the bulk of matter in the
universe is not the normal electron-nucleon staff but something unknown
which may be made of not yet discovered elementary particles, or
of some field excitations at
astronomically large scales which are stable by topological
reasons, or maybe even something else. This unknown form of matter
contributes more than 90\% into total mass (or better to say, energy)
density in the universe. This matter is usually called dark matter.
The name reflects the fact that it is not luminous and maybe also the
level of our knowledge of the subject. A better name is invisible
matter since it neither emits nor absorbs light. The only
observed manifestations
of this matter are gravitational effects. There
are also some theoretical arguments in favor of
existence of dark matter but,
though very persuasive, they are not absolutely compelling.
It is the biggest challenge in physics of
this and may be of the next century to directly observe this new (if
it is indeed new) form of matter and to study its properties.

In these lectures I will address he following questions:
\begin{enumerate}
\item{} Is there indeed dark (invisible) matter and why do we think
that it is so?
\item{} Could all dark matter be baryonic? If not all, could there still
be some invisible baryons?
\item{} If not baryonic, what is it? There are two ways to address this
question: one is to use the elementary particle theory for possible
candidates for the constituents of dark matter and the other is to use
the theory of large scale structure formation in the universe and the
astronomical observations to constraint the properties of dark matter
particles.

\end{enumerate}

In the next Section the necessary cosmological background is briefly
presented.
In sec. 3 the luminous matter is discussed.
Nucleosynthesis constraints on the
amount of baryons in the universe are presented in Sec. 4.
Astronomical data indicating gravitational action of invisible
matter are considered in Sec. 5.
In sec. 6 the so-called baryonic crisis created by a too large
an amount of hot gas observed in rich galactic clusters is discussed.
Prediction of inflationary cosmology on the value of $\Omega$ is
given in sec. 7.
In sec. 8 the summary of measurements of $\Omega$ for different
forms of matter is presented. A brief
discussion of possible cosmic relics which may exist in the universe
and contribute to the dark matter is given in sec. 9.
In sec. 10 a critical analysis of the basic assumptions of
the theory of large scale structure formation is presented.
A brief introduction to the basics of the theory of structure
formation is given in sec. 11. Evolution of density perturbations in
the universe dominated by different forms of dark matter is
described in sec. 12.
In Conclusion some speculations on
the best bet about the form of dark matter is presented
and possible experiments and theoretical problems,
which may shed light on the nature of dark matter, are discussed.
I will not review the existing direct experimental searches of
dark matter particles in low background experiments. It is done by
L.Mosca in this School.

\section{Cosmological Background }

The source of gravitational field is the energy-momentum tensor
$T_{\mu\nu}$. In homogeneous cosmological models it is presented
in the ideal liquid form which in the rest frame of matter can be
written as:
\beq{
T_{\mu\nu} = diag (\rho,-p,-p,-p)
}\label{tmunu}
\eeq
where $\rho$ and $p$ are respectively the energy and pressure density.
It is usually assumed that at a later stage the universe is dominated
by nonrelativistic matter when $p$ may be neglected and
the energy density of nonrelativistic matter, $\rho = \rho_m$, is
practically equal to the mass density. Indeed as a function of the
scale factor $a(t)$, which describes the universe expansion,
the energy density of nonrelativistic matter goes down as
$\rho_m \sim a^{-3}$, while that of relativistic one as
$\rho_{rel} \sim a^{-4}$.  The late dominance of nonrelativistic
matter may be not true in models with long-lived decaying particles
producing relativistic species but in the standard cosmology it
is fulfilled with a very good precision. Apart from
contributions to $T_{\mu\nu}$ from normal
relativistic or nonrelativistic matter there may be
also a rather mysterious contribution from vacuum energy with
$T_{\mu\nu} = \rho_{vac} g_{\mu\nu}$ where $g_{\mu\nu}$ is the
metric tensor. In contrast to the energy density of matter
$\rho_{vac}$ stays constant in the course of the universe
evolution. Not long ago a nonzero $\rho_{vac}$ was considered as
rather exotic by majority of astronomers and was usually neglected
but now with accumulation of new data the attitude is changing.

Roughly speaking the matter in the universe can be divided into
three categories: 1) Visible or luminous matter which is observed
either by emitted or absorbed light. Almost surely this is the normal
baryonic staff. 2) Invisible normal baryonic matter. We can make
a conclusion about its amount considering primordial
nucleosynthesis. 3) Invisible nonbaryonic matter. This one is observed
only by its gravitational action. It can also be divided
according to the spatial
distribution: clustered or uniform.
We can "feel" the former by observing the motion of
the visible tracers like gas around galaxies or galactic
satellites while the latter can be observed only by cosmological
effects and hence the observations are more complicated and
less secure. In particular vacuum energy belongs to
the last case.

The magnitude of mass or energy density in the universe,
$\rho$, is usually presented in terms of the dimensionless parameter
\beq{
\Omega = \rho / \rho_c
}\label{omega}
\eeq
where $\rho_c$ is the so called critical or closure density,
\beq{
\rho_c = { 3H^2 m_{Pl}^2 \over 8\pi } = 1.88 h_{100}^2 \times
10^{-29} \,g/cm^3 = 10.5 h_{100}^2 \,KeV/cm^3
}\label{rhoc}
\eeq
Here $m_{Pl} = 1.22\times 10^{19}$GeV is the Planck mass (the
Newtonian gravitational constant is $G_N = m_{Pl}^{-2}$)
and $h_{100}$ is the dimensionless Hubble parameter,
\beq{
H = 100 h_{100}\,km/sec/Mpc
}\label{H}
\eeq
The value of $H$ is very poorly known, $ 0.4 < h_{100} < 1 $. The
universe age, as we see below, favors lower values of $H$ but the recent
data indicate that $ h_{100} = 0.7 - 0.8 $ or maybe even
higher\cite{hubble}.

Integrating the Einstein equation:
\beq{
H^2 \equiv \left( {\dot a \over a} \right)^2 =
{8\pi \rho \over 3 m_{Pl}^2 } - {k \over a^2},
}\label{h2}
\eeq
which governs the expansion of the universe,
one can express the universe age
through the present day value of the Hubble constant and the fractions
of energy density of different forms of matter (for more detail see
e.g. my lecture in the previous ITEP school \cite{ad1}
or the textbooks \cite{zn,sw,kt,dzs}):
\beq{
t_u = {1 \over H} \int_0^1 {dx \over \sqrt{ 1 - \Omega_{tot} +
\Omega_{m} x^{-1} + \Omega_{rel} x^{-2} + \Omega_{vac}x^2 }}
}\label{tu}
\eeq
where $\Omega_{m}$, $\Omega_{rel}$, and $\Omega_{vac}$
correspond respectively to the energy density of nonrelativistic
matter, relativistic matter, and to the vacuum energy density
(or, what is the same, to the cosmological constant);
$\Omega_{tot} = \Omega_{m} + \Omega_{rel} + \Omega_{vac}$,
and $1/H = 9.8 h_{100}^{-1}\times 10^9$yr.

The universe age is estimated by two different methods: by nuclear
chronology which uses measurements of the ratios of the long-lived
isotopes, $^{187}Re / ^{232}Th$ or $ ^{238}U / ^{235}U$, and by
the estimated age of old stellar clusters. Both methods give the age
in the range of (12 - 20) Gyr. The recent analysis gives
$t_u \approx 18$ Gyr (for the review and the list of references see
e.g. paper \cite{vdb}).

If the universe is dominated by nonrelativistic matter, as is usually
assumed, i.e. $\Omega_{tot} = \Omega_{m}$, then the universe age is
approximately given by the expression
$t_u \approx 9.8 h_{100}^{-1} / (1 + \sqrt \Omega_m /2 )$ Gyr. Even with
$\Omega = 0$ the big universe age and large $H$ are inconsistent. The
inconsistency becomes stronger for $\Omega_{tot} = 1$ predicted by
inflationary models. The observed tendency to large values of $H$ and
$t_u$ presents a very strong argument in favor of nonvanishing vacuum
energy. For $\Omega_{tot} = \Omega_{m} + \Omega_{vac} = 1$ eq.(\ref{tu})
gives
\beq{
t_u = {2 \over 3H \sqrt \Omega_{vac}} \ln { 1 + \sqrt \Omega_{vac} \over
  \sqrt { 1 -  \Omega_{vac} }}
}\label{tuvac}
\eeq
To get, let us say, $t_u = 14$ Gyr for $h_{100} = 0.75$ we need
$\Omega_{vac}\approx 0.8$. If this is true the bulk of energy in the
universe is just the energy of empty space, of vacuum. Unfortunately
our understanding of vacuum energy is very poor. Any reasonable
estimate of it gives the result which is some 50-100 orders of magnitude
higher than the observational limit $\rho_{vac} < 10^{-47}\,GeV^4$
(for the review see refs. \cite{sw2,ad2}). For example there are
contributions from quark and gluon condensates into $\rho_{vac}$
which are well established in QCD and which are
about $10^{-4}\, GeV^4$.
So there must exist a contribution into vacuum energy from
something not related
to quarks and gluons but with exactly the same magnitude and
the opposite sign. It is hard to
imagine an accidental cancellation with such an accuracy but no
dynamical mechanism has yet been found. If something is very small,
one would naturally expect this to be exactly zero and this is a
general attitude to vacuum energy. This is one of the reasons why
models with a nonzero cosmological constant were not seriously
considered by the establishment. Another reason for that is an
unnatural coincidence of the present-day values of vacuum
energy and the critical energy density of the universe. As we
mentioned above they are pretty close to each other. However
the critical energy density is decreasing with the universe
age as $1/t^2$ while $\rho_{vac}$ remains constant (at least in the
standard model). The new astronomical observations
changed the attitude to cosmological
constant and now it is considered more seriously. Moreover, as we
see below, models with nonzero vacuum energy have some advantages in
the description of the structure formation in the universe. To
my mind the best possibility in solving the cosmological
constant problem is the adjustment mechanism \cite{ad3,ad4} which
ensures the cancellation of vacuum energy by an action of a new
field coupled to gravity or curvature so that its condensate
cancels out any initial $\rho_{vac}$. This cancellation is
generically not complete and a noncompensated amount of
$\rho_{vac}$ is always of the order of $\rho_c$. In such cosmological
models both vacuum energy and the energy of the new field are
essential at any stage of the universe evolution (in contrast to
the models with normal time independent vacuum energy) and this
might have an impact on primordial nucleosynthesis, structure formation,
etc. The progress here is inhibited by an absence of a consistent
(even toy) model.

\section{Luminous Matter }

The amount of luminous matter in the universe is estimated in the
following way. The flux of light, $F$, from different sources is
measured on the Earth and the luminosity of an individual source
is found by the relation:
\beq{
L_i = 4\pi F_i l^2_i
}\label{l}
\eeq
where $l_i$ is the distance to the source. Then one may estimate the
average mass density of luminous matter as
$\rho_{lum} = \sum_i L_i (M_i/L_i) /V $ where $V$ is a (large) volume
over which the averaging is done. Usually the value
$M/L =5M_\odot / L_\odot$, which is typical for the stellar
(galactic) matter, is substituted for the ratio $M_i / L_i$
(here $M_\odot = 2\times 10^{33}\,g$ and
$ L_\odot =4\times 10^{33}\,erg/sec$
are respectively the solar mass and luminosity).
The measured flux permits to find the luminosity density as
(see e.g. ref. \cite{bt}):
\beq{
\lambda = \sum_i L_i  /V =
(2 \pm 0.6) \times 10^8 h_{100} L_\odot \,Mpc^{-3}
}\label{lambda}
\eeq
where $1 \, Mpc = 3.1\times 10^{24} \,cm$.
The result is proportional to the Hubble constant because the
distance $l$ is determined  by the Hubble flow, $v=Hl$ and
correspondingly $l^2/V \sim h_{100}$. The mass density of
luminous matter is given by $\rho_{lum} = \lambda (M/L) =
5(\lambda /L_\odot) M_\odot = (0.7\pm 0.2)\times 10^{-31}
h_{100}\,g/cm^3$.

Using these numbers we find for the relative amount of the luminous
matter:
\beq{
\Omega_{lum} = \rho_{lum} / \rho_c = (0.35 \pm 0.10)\% h_{100}^{-1}
}\label{omegalum}
\eeq
This is by far smaller than the amount of nonluminous matter.

\section{ Primordial Nucleosynthesis and the Net Amount of Baryons }

Though some (or maybe even the bulk) of baryons are invisible there
is a way to find their total number density with a relatively good
precision. This is based on consideration of primordial
nucleosynthesis of light elements such as deuterium, helium-3,
helium-4, and lithium-7. Their calculated abundance are sensitive
to the cosmic baryon number density or, better to say, to the
baryon-to-photon ratio, $\eta_{10} = 10^{10} n_B / n_\gamma
= 10^{10} \eta$.
The starting point of the nucleosynthesis is the fixation of the
proton-to-neutron ratio. It is determined by the competition between
the weak interaction reactions, $pe^- \leftrightarrow n\nu_e$ and
$p \bar \nu_e \leftrightarrow n e^+ $, and the universe expansion
rate. To the moment when these reactions become ineffective the
(n/p)-ratio freezes approximately at 1/6. It takes place roughly
at $T=0.65\,MeV$. Later on at $T\approx 0.065\, MeV$,
when the formation of light nuclei begins, this ratio
drops down to 1/7 because of neutron decay. The magnitude
of $n/p$-ratio does not depend upon $\eta$ because the corresponding
reactions are linear with respect to baryons and so their rate,
$\dot N/N$, is
independent of the baryon number density. Note that $n/p$-ratio
is rather sensitive to the number of different particle species in the
primeval plasma and this permits in particular to put a bound
on the number of different massless ($m<1 \,MeV$) neutrino
species. For the up-to-date analysis of this bound see
refs.\cite{ost,os,hss}.

Since $\eta \approx 3\times 10^{-10}$ is a very small number, the
amount of the produced nuclei is tiny even at the temperatures
which are smaller than their binding energy. For
example the number density of deuterons is determined by
chemical equilibrium and is equal to
\beq{
n_D = n_n n_p e^{B_D /T} \left( {m_D T \over 2\pi} \right)^{3/2}
     \left( {2\pi \over m_N} \right)^3
}\label{nd}
\eeq
where $B_D = 2.224 \,MeV$ is the deuterium binding energy. One can see
that $n_D$ becomes comparable to $n_n$ at the temperature:
\beq{
T_D = { 0.065 \,MeV \over 1 - 0.03 \ln \eta_{10}}
}\label{td }
\eeq
Above this temperature the number density of deuterons in the cosmic
plasma is negligible and correspondingly the formation of other nuclei,
which proceeds through collisions with deuterium is suppressed. This is
the so called "deuterium bottleneck". But as soon as $T_D$ is reached,
the nucleosynthesis goes very quickly and practically all the neutrons,
which existed in the cosmic plasma at that time, are captured into
helium-4. The latter has the largest binding energy,
$B_{^4 He} = 28.3 \, MeV$ and so in equilibrium
its abundance should be the largest. Its mass fraction, $Y(^4 He )$,
is determined predominantly by the $(n/p)$-ratio and is equal to
$2(n/p)/[1+(n/p)] \approx  25\%$.

The time moment corresponding to $T=T_D$ is determined by the well known
relation:
\beq{
t_D = 0.16 \left({3.37 \over g_*}\right)^{1/2} \left({m_{Pl}
    \over T^2}\right) = 310\,sec \left({3.37 \over g_*}\right)^{1/2}
   \left( 1 - 0.03 \ln \eta_{10} \right)^{ 2}
}\label{t}
\eeq
This relation is obtained by equating the expression for the critical
energy density of relativistic matter
\beq{
\rho_c = 3m_{Pl}^2/32\pi t^2
}\label{rhoc1}
\eeq
(see eq.(\ref{rhoc}) with $H=1/2t$) and the energy density of
equilibrium relativistic plasma with temperature $T$
\beq{
\rho_T = \pi^2 g_* T^4 /30
}\label{rhot}
\eeq
Here $g_*$ is the number of relativistic species in the plasma. For
a photon and three types of neutrinos with equal temperature
$g_*=7.25$. After electron-positron annihilation the temperature of
photons became 1.4 larger than that of neutrinos and $g_*=3.37$.
One sees that with
a larger $\eta$ the time $t_D$ is getting smaller and the number of
not-yet-decayed neutrons is larger, $n_n \sim \exp (-t_D / \tau_n )$,
where $\tau_n = 890\,sec$. Thus $Y(^4 He)$ is a rising function of
$\eta$ and its variation is relatively weak, with $\eta_{10}$
changing from 1 to 10
the mass fraction of helium changes by approximately
10\%. Still since $Y(^4 He)$ is rather accurately known it can be
used as a serious indicator of $\eta$ and correspondingly of the
{\bf total} baryonic mass density.

In contrast to $^4 He$ the fraction of deuterium is a strong and
a decreasing function of $\eta$. Physically it is easy to understand:
more baryons are in the plasma, more efficient are the processes of
disappearance of deuterium through
formation of heavier nuclei with larger binding energies, like e.g.
$D+p \rightarrow ^3 He + \gamma$ and
$^3 He+n \rightarrow ^4 He + \gamma$.
The destruction of deuterium in the first
process is governed by the equation:
\beq{
\dot n_D = - \sigma v n_D n_p
}\label{dotnd }
\eeq
where $\sigma$ is the cross-section of the reaction and $v$ is the
velocity of the colliding particles. We assumed that the temperature
is low enough so that the inverse process is Boltsmann suppressed and
is not effective.

Integrating this equation under assumption that $\eta_p = n_p/n_\gamma
\approx \eta = const$ we find:
\beq{
n_D = n_{Di} \exp [-0.06 \eta (3.77/g_*)^{1/2} \sigma v m_{Pl} T_i]
}\label{ndi}
\eeq
where $n_{Di}$ and $T_i$ are the initial values of the number density
and temperature which should be taken at the moment when the
production of deuterium  by photo-dissociation of
heavier nuclei effectively stops.
Eq. (\ref{ndi}) illustrates the made above statement
about the dependence of deuterium abundance on $\eta$.
$^3 He$ has the similar dependence on $\eta$ but somewhat less strong.
The net output of these elements is much smaller than that of $^4 He$,
$D/H \approx 10^{-4}$ and $^3He /H \approx(1-2)\times 10^{-5}$ by number.

The fraction of $^7 Li$ depends on $\eta$ nonmonotonically
having a minimum at $\eta_{10} \approx 3$. This more complicated
dependence is connected with a competition of different production
channels.

Though the abundance of $D$, $^3 He$, and $^7 Li$ are not so well
known as that of $^4 He$, their strong dependence on $\eta$ also
permits to use them as indicators of the total baryonic mass density
in the universe.

Of course the calculations of the light element abundance are not
done in this naive way. We use it here only for the qualitative
description of the phenomenon. The calculations are based on the
numerical integration of the
complete set of kinetic equations with the experimentally measured
nuclear cross-sections. With the fixed values of the parameters
the theoretical results are very accurate but the direct comparison
with observation is difficult not only because of the observational
uncertainties but also because of complicated evolutionary effects.
Due to the latter the light element abundance in the present-day
universe may be quite different from the primordial ones. It is
relatively simple with $^4 He$ because it can only be produced but
not destroyed in the course of the universe evolution.
The production of helium-4 in stars
is accompanied by the production of oxygen, nitrogen, and carbon
(they are called metals by astronomers). Because of that
the observations of primordial helium are done in the
regions of the sky least
contaminated by these heavier elements.
Moreover the data on helium-4 and the heavy elements are linearly
extrapolated to the so called zero-metallicity state. In this way a
rather safe upper bound on $Y(^4 He)$ can be obtained.

The evolutionary behavior of
other light elements is more complicated, they
can be both destroyed and produced in stellar processes and the
comparison of the theory with observations is difficult. Still only
two years ago the agreement between the the theory and observations
was believed to be excellent. In a sense this is true because the
theory predicts the abundance of different light elements which
span the region between 25\% for $^4He$ down to $10^{-10}$ for
$^7 Li$ and this is definitely confirmed by the data. The problem
appeared at the level of more refined conclusions. The old
data for all elements: $^4 He$, $D$, $^3He$, and $^7 Li$,
were in agreement with the theory for the common value of
$\eta_{10} = 3.5\pm 1$. The best fit to the data was obtained with
the three known massless neutrinos which was also an argument in
favor of the overall agreement. With this value of $\eta$ the
baryonic mass fraction in the universe is
\beq{
\Omega_B = 4\times 10^{-3} \eta_{10} h_{100}^{-2} =
(1.4 \pm 0.4 )\% h_{100}^{-2}
}\label{omegab1}
\eeq
In this case
$\Omega_B /\Omega_{lum} = (2-7) h_{100}^{-1} $ and the baryons in
the universe are mostly invisible.

A new measurement of deuterium abundance \cite{crw,sch} at high redshift
$z=2.9$ showed a surprisingly high value,
$D/H  = (1.9-2.5)\times 10^{-4}$. It was done by the measurement of
absorption of light from
a distant quasar in a Lyman alpha cloud. The
deuterium in this cloud did not suffer from evolution and so its
abundance should be equal to the primordial one. This value is
an order of magnitude bigger than the value found in interstellar
medium. This would request $\eta$ approximately
three times smaller than the
previously accepted and, if so, practically all baryons in the universe
are visible. The analysis of the new data on helium-4 \cite{os} also
suggests a smaller $\eta$.
However another group \cite{tf} reported much smaller
value, $D/H = (1-2)\times 10^{-5}$. So the problem remains unsettled.
A smaller value of $\eta$ creates problems with the abundance of
$^3 He$ which in this case
should be considerably larger than that observed in our
neighbourhood. For the analysis of the new data and evolutionary
effects one can address the recent papers \cite{os,ost,cst}. In
accordance with the analysis of helium-4 made in
ref. \cite{os} the best value for
the number of neutrino species is $ N_\nu = 2.2 \pm 0.27 \pm 0.42$
if the old value $\eta_{10} = 3.6$ is assumed. Though the
central value of $N_\nu$ is below three the error bars are generous
enough to be consistent with the standard three neutrinos.

So we have two interesting possibilities which can be resolved
with an improved accuracy. The first is that $\eta$ is close
to the old value so that there are invisible baryons and $N_\nu <3$.
It may mean in particular that $\nu_\tau$ is heavy and unstable.
Another more conservative possibility would be realized if
$D/H$ is high, $\eta$ is small and thus all baryons are visible,
and $N_\nu =3$. Small $\eta$ may imply a problem with
the observed $^3 He$ but evolutionary effects
could be significant (see ref. \cite{os}). A large universe age looks
favorable from the point of view of primordial deuterium destruction.
This can explain a large amount of primordial $D/H$ observed in Lyman
alpha clouds and a small amount of it in interstellar medium nearby.
The large $t_U$ in
turn is possible only with nonzero vacuum energy. If an adjustment
mechanism for its cancellation is operating then the noncompensated
amount was essential during all stages of the universe evolution and
this might have an unknown impact on the primordial nucleosynthesis.
However if we forget about this rather uncertain possibility,
the conclusion that the primordial nucleosynthesis does not
permit the mass fraction of baryons to be larger than a few per cent
seems to be sound enough.

\section{Gravitating Invisible Matter }

The observation that there is more gravitating matter in the universe
than is directly seen was made more than half a century ago
by Oort \cite{oo} for the Galaxy and by Zwicky \cite{zw} for Coma
cluster. Virial considerations showed that the visible matter alone
cannot account for the observed velocities. To the present day
a very rich data were accumulated which strongly support the conjecture
that there is a large amount of invisible matter in the universe.
Velocities of HI gas clouds around galaxies are measured by 21 cm line
for hundreds of galaxies up to distances $r_{max} = 30 \,Kpc$. All
this curves becomes flat, $v \rightarrow const$, for $r$ outside the
luminous centre, $r>r_{gal} = 10\,Kpc$. The equilibrium velocities are
determined by the relation
\beq{
{v^2 \over r } = {G_N M(r) \over r^2}
}\label{v2r }
\eeq
where $G_N=m_{Pl}^{-2} $ is the Newtonian gravitational constant
and $M(r)$ is the mass inside radius $r$.
This equation implies that for the mass
confined inside $r_{gal}$ and for $r>r_{gal}$ the velocity
goes down with the radius as
$1/\sqrt r$, while for $v$ tending to a constant, $M(r) \sim r$. Since
$r_{max}$ is considerably larger than the luminous galactic radius
the amount of invisible matter in galaxies should be much larger that
that of the visible one.

The law $M(r)\sim r$ implies that the mass density $\rho (r)$ falls
down as $1/r^2$. This takes place for the isothermal selfgravitating
gas as follows from the equations:
\beq{
\Delta \phi = 4\pi G_N \rho
}\label{deltaphi}
\eeq
and
\beq{
\rho = \rho_0 \exp (-\phi /T)
}\label{rho0 }
\eeq

There is a very important question how far, that is to what maximum
value of $r$, the law $M(r)\sim r$ or $v= const$ is valid. The
measurement of HI gas velocities which can be done up to 30 Kpc
does not show any cutoff and for large spirals give
$M_{DM} / M_{lum} = 3-5$. Measurements of velocities of satellite
galaxies around the Milky Way and other large spirals can be extended
up to 200 Kpc and give $M_{DM} / M_{lum} = 12-15$. This gives
$\Omega_{spirals} \approx 0.09 h_{100}^{-1}$.  Whether dark halos
go beyond this distance is unknown. No cut-off is observed
with all existing observations. Elliptical galaxies where dark matter
is traced up to 100 Kpc by hot X-ray gas
give $M_{DM} / M_{lum} \geq 10 $. An last but not the least (though
less secure) the clusters of galaxies give $M_{DM} / M_{lum} = 50-100$.

The determinations of the gravitating masses by rotational velocities,
considered above, are valid for gravitationally bound systems and
applicable to masses at relatively small scales, at most of cluster of
galaxies.There is another way to determine gravitating masses which can
be applied to
considerably larger scales. It is based on the measurements of the
large scale flows. The latter are the so called peculiar velocities i.e.
the velocities of different galaxies with
respect to the general Hubble expansion. By assumption these
velocities are induced by the density contrast,
$\delta = ( \rho - \rho_0 )/ \rho_0$, where $\rho_0$ is the
homogeneous cosmological energy density. In other words peculiar
velocities are proportional to the gravitational acceleration created
by the contrast of the gravitational potential, $\delta \phi$. The theory of
this phenomenon is simplified because at large scales $\delta \ll 1$
and the linear approximation is valid.

{}From  the continuity equation, $\partial_t \rho +
\vec \partial (\rho \vec v) = 0$, one can find in the linear regime:
\beq{
\vec \nabla \vec v = -  \partial_t \delta
}\label{nablav}
\eeq
where $a(t)$ is the cosmological scale factor.

For sufficiently large wave lengths (larger than the Jeans wave length,
see below sec. 11) the behaviour of $\delta$ at matter dominated stage
is governed by the equation (see e.g.
\cite{peeb,efst}):
\beq{
\ddot \delta + 2 (\dot a/a) \dot \delta -
4\pi G_N \Omega \rho_0 \delta = 0
}\label{ddotdelta}
\eeq
This equation can be easily solved in two limiting cases:
1) $\Omega = 1$ for which the rising (gravitationally unstable) mode
goes as $\delta \sim a(t) \sim t^{2/3} $ and 2) $\Omega = 0$ for which
$\delta = const$. In the intermediate case, $0<\Omega <1$, the solution
can be approximately presented as \cite{peeb}:
\beq{
(\dot \delta /\delta ) = \Omega^{0.6} (\dot a / a) \equiv H\Omega^{0.6}
}\label{dotdelta}
\eeq
So from eq. (\ref{nablav}) follows
\beq{
\vec \nabla \vec v = - H\Omega^{0.6} \left({\delta \rho \over \rho}
\right)_{tot} = -{H\over b} \Omega^{0.6} \left({\delta \rho \over \rho}
\right)_{vis}
}\label{nablav1}
\eeq
where the biasing factor $b$ is introduced to relate the visible density
contrast to the total one. Essential assumptions which are made
in the derivation of this equation are the linearity
i.e $\delta_{tot} = \delta_{vis} /b$  and the scale independence of this
relation. It is also assumed that the visible density contrast is given
by the excess in galaxy number count $\delta n /n$. All these
assumptions may be incorrect to a larger or smaller extent.

It is noteworthy that this equation does not contain the ambiguity
connected with the cosmological distance scale or (what is the same) with
the value of the Hubble constant $H$. Indeed the integration of the
equation gives $\int Hdr \sim v_H$ where $v_H$ is the velocity of the
Hubble expansion and is directly measurable. Note also that
our derivation of eq.(\ref{nablav}) is oversimplified and can be used
for illustrative purposes only; for a more rigorous approach see e.g.
refs.\cite{kt,peeb}.

The velocities of galaxies are measured by the red-shift and so only the
component in the direction to the observer can be found. This
extra problem can be solved if the velocity field is rotationless,
$ \vec \nabla \times \vec v = 0$ \cite{bd}.
It is true for pure gravitational interactions for which
an initially curlless velocity field remains curlless. Thus the velocity
field can be expressed through a single scalar function, velocity
potential, and this permits to determine $\vec \nabla \vec v$ from
observations and to find $\Omega$ from eq. (\ref{nablav1}). The recent
analysis \cite{hdc} of peculiar velocities of about 3000 galaxies gives
\beq{
\Omega^{0.6}/b = 0.74 \pm 0.13
}\label{omegab}
\eeq
So if $\Omega = 1$ then $b=1.35\pm 0.23$ or if $b=1$
then $\Omega =0.61\pm 0.18$.
If $\Omega <0.3$ one has to introduce antibiasing, $b<1$.
Though it is not absolutely excluded, the physics of that is
rather unclear.
For the review of this method and possible errors one can address to the
paper \cite{dek}. It is argued that the lower bound on $\Omega $
independent of biasing is $\Omega > 0.3$. This conclusion however is
based on the assumption of vanishing cosmological constant. It is
worthwhile to redo the analysis without this assumption. Some other
potential reasons that may invalidate the above conclusion are a possible
dependence of $b$ on scale and a relatively small
size of the analyzed sample of galaxies. Another reason which
could also lead to the
violation of the statement about large $\Omega$ is a possible
nongravitational contribution to the large scale flow. If for example
the large observed dipole anisotropy of the cosmic microwave background
(cmb) is not induced by the gravitational acceleration but has its own
(intrinsic) origin like isotemperature fluctuations
with very large (superhorizon) wave length, the
results based on the measurement of the large scale flow may be invalid.
Typically this kind of mechanism would create the quadrupole $q$
and higher multipole anisotropy
of cmb of the same order as the dipole ($d$) while observationally
$d\approx 2\times 10^{-3}$ and $q= (a few)\times 10^{-6}$. This problem
is analyzed in the recent paper \cite{lp} where it is argued that it
is still possible to get a large dipole while all other multipoles
would be small. Note that the model of baryon island universe \cite{dk},
where the baryon asymmetry was generated only inside a spherical bubble
with the present-day size corresponding to $z=5-10$,
provides $q\approx d^2$ and very small other multipoles. The dipole and
other asymmetries in this model are connected with the shift in our
position with respect to the geometrical center of the island. The
smallness of the dipole means that we live almost in the Center of the
Universe. This is not very natural but rather appealing with regard of
our vanity.

\section{Baryonic Crisis}

Recent observations \cite{wnef,bc} showed a surprisingly large fraction
of baryons in rich clusters of galaxies. In the first paper \cite{wnef},
which drew attention to the problem, the mass of the hot gas in the
Coma cluster was estimated by the intensity of
X-ray emission measured by Rossat
satellite, $M_{gas} = (5.45 \pm 1) \times 10^{13} h_{100}^{-5/2}
M_\odot $. It dominates the mass of matter contained in the stars,
$M_{star} = (1\pm 0.2)\times 10^{13} h_{100}^{-1} M_\odot $.
The total
mass of the cluster was found by two methods: by the virial consideration
and by the condition of the thermal hydrostatic equilibrium in the gas;
both ways give the close results:
$M_{total} = (6.7\pm 1)\times 10^{14} h_{100}^{-1} M_\odot $. Thus the
ratio of the mass contained in baryons to the total mass of the
cluster is
\beq{
r_B \equiv {M_{bar} \over M_{tot} } = (0.09 \pm 0.05) h_{100}^{-3/2}
}\label{rb1}
\eeq
Similar results have been obtained in ref.\cite{bc} from the analysis
of 5 clusters (Coma plus four others):
\beq{
r_B > {M_{gas} \over M_{tot} } = (0.03 - 0.08) h_{100}^{-3/2}
}\label{rb2}
\eeq
As has been argued in ref. \cite{sf} these results give safe lower bounds
on $r_B$ and the real numbers should be noticeably higher.

Now if we assume that the mass contained in the clusters represents a
fair sample of the total mass in the universe then these data
together with the nucleosynthesis constraint on $\Omega_B$ put the
strong upper bound on the energy density in the universe:
\beq{
\Omega_{cluster} < 0.05 \eta_{10} h_{100}^{-1/2} (0.1/r_B )
}\label{omegac}
\eeq
This bound is evidently valid for the clustered matter and is not
applicable for the uniformly distributed one.

The authors of ref. \cite{wnef} have discussed several different
explanations of the phenomenon like nonstandard nucleosynthesis,
nongravitational processes in the structure formation, possible mistakes
in the standard technique in the estimation of the cluster mass, and
an open universe model with $\rho_{tot}\ll \rho_c$. The last possibility
is disfavored by the standard inflationary scenario (see below).
To my mind these data give a good indication to nonzero vacuum energy,
$\Omega_{vac} \approx 0.8$.
It is uniformly distributed and for the typical cluster size about
1 Mpc does not
contribute much inside this radius. The nonzero value
of vacuum energy is
rather natural by the reasons mentioned above. Another evident
possibility is a large contribution from the so called hot dark matter
which may be also nonclustered at these scales. This however seems to
be excluded because the total amount of hot dark matter in the universe
cannot exceed 30\% at least in the models with scale-free
spectrum of initial perturbations (see below Sec. 12)

\section{Inflationary Omega}

Aesthetically $\Omega = 1$ is definitely the most
attractive possibility. It
is the only value of $\Omega$ which is not changing with time in the
course of the universe evolution. In the universe dominated by
any normal matter all other values of $\Omega$ run
away from 1. To see this let us rewrite
eq.(\ref{h2}), using definitions (\ref{omega}) and (\ref{rhoc}),
in the following way:
\beq{
\Omega^{-1} = 1 - {C \over \rho a^2}
}\label{omega-1}
\eeq
where $C=3km_{Pl}^2 / 8\pi = const$. For nonrelativistic matter with
negligible pressure $\rho \sim 1/a^3$ and for relativistic one with
the pressure $p=\rho /3$, $\rho \sim 1/a^4$. In the general case the
behaviour of $\rho$ is governed by the covariant energy conservation:
\beq{
\dot \rho = -3H(\rho + p )
}\label{ }
\eeq
and for normal matter with positive pressure $\rho < 1/a^3$ when
$a \rightarrow \infty$. Thus for an open universe ($C<0$) dominated by
nonrelativistic matter $\Omega\sim 1/a \rightarrow 0$ when
$a\rightarrow \infty$ and for the closed universe ($C>0$)
$\Omega \rightarrow \infty$.

For vacuum energy $p=-\rho$ and so $\dot \rho = 0$ and
$\rho a^2 \rightarrow \infty$. Correspondingly $\Omega \rightarrow 1$
with rising $a$. It would be very interesting if an adjustment mechanism
is found which would lead to the equation of state such that
$\rho a^2 =const$ which in turn would lead to
$\Omega$ unchanging in the course of expansion.

Assuming that at the present day $\Omega$ is somewhere between 0.1
and 2, we see that the law (\ref{omega-1}) implies $(\Omega -1)=
O(10^{-15})$ at the nucleosynthesis epoch. At earlier stages the
finetuning should be even stronger. This is the well known flatness
problem which was beautifully solved by the inflationary
proposal \cite{ag,asl}. Typically in most inflationary models the universe
is dominated by the vacuum-like energy so that $\rho = const$ and
$a \sim \exp (H_It)$. Correspondingly $(\Omega_{infl}^{-1} -1)
\sim \exp(-2H_I\tau) (\Omega_{i}^{-1} -1)$ where $\tau$ is the
duration of inflationary stage, $H_I$ is the Hubble parameter
during inflation, $\Omega_{infl}$ is the value
of $\Omega$ at the end of inflation, and $\Omega_i$ is the initial
(preinflationary) value of $\Omega$. When inflation is over, $\Omega$
runs away from 1 and at the present day reaches the value:
\beq{
\Omega_0^{-1} -1 = (\Omega_{infl}^{-1} -1)(z_{infl} +1)^{-2}
  (\rho_{infl}/\rho_0 ) \approx (\Omega_{infl}^{-1} -1) (T_R/T_0)^2
}\label{omega0-1}
\eeq
where subzero means the present-day values, $(z_{infl} +1)=(a_0/a_{infl})
\approx (T_R/T_0)$ is the red-shift corresponding to the end of
inflationary era, $T_R$ is the temperature of the universe heating at
the end of inflation, and $T_0 = 2.7\,K$ is the
present-day temperature of the cosmic
microwave background radiation. We assumed that the energy density at
the end of inflation is $\rho_{infl} = (\pi^2 /30)g^R_*AT^4$
where $g^R_*$
is the number of relativistic species at $T=T_R$ and $A>1$ is the
coefficient which accounts for the slow process of heating due to the
weak inflaton coupling. The last equation
in (\ref{omega0-1}) was obtained under
assumption $Ag^R_*(\rho_{cmb})/\rho_0 \approx 1$. The necessary duration
of inflationary stage is approximately equal to $H_I\tau \approx
\ln (T_R/T_0) = 52 + \ln (T_R /10^{10}\,GeV)$. This gives $\Omega_0^{-1}
-1 = O(1)$. Even a slightly longer inflation would result in
$\mid \Omega_0^{-1} -1 \mid \ll 1$ and a slightly shorter one would
be in a gross disagreement with observations.
Typically inflationary models give
$H_I\tau \gg 1$ which results in $\mid \Omega_0^{-1} -1 \mid
\leq 1\pm 10^{-4}$. The deviation from 1 in the r.h.s. of
this relation is connected to the density inhomogeneity at the present
horizon scale. So $\Omega = 1$ is considered as a robust prediction
of inflationary cosmology.

Recently with accumulation of new astronomical data which may
request $\Omega <1$ there appeared models where a possibility of having
$\Omega \neq 1$ in inflationary frameworks was advocated
\cite{bgt}. All these models request a tuning of inflationary
and postinflationary stages in such a way that $\exp(2H\tau) =
(T_R/T_0)^2 (\Omega_0^{-1} -1)^{-1}$. This condition looks very strange
because physically these epochs are not related. The
condition of small density perturbations which is not easy to realize
in these models imposes an extra restriction making these models even
less natural.

Thus even if $\Omega \neq 1$ is in principle possible in inflationary
models the flatness problem in such models is similar to that in the
old Friedman cosmology. Still it is very difficult (if possible) to
create our universe without inflationary stage and if it is indeed
proven that $\Omega_{matter} < 1$ a much better solution which naturally
agrees with inflation is a nonzero vacuum energy such that
$\Omega_{tot} =\Omega_{matter} + \Omega_{vac} = 1$. On the other hand
if we observe that $\Omega_{tot} \neq 1$ it would be a strong argument
against inflation (I have to admit that this opinion is not shared by
everybody).

\section{Intermediate Summary}

Let us summarized what is known about energy density of different
forms of matter in the universe.
\begin{enumerate}
\item{}The luminous matter which is undoubtfully made of the usual
particles, protons, neutrons, and electrons, contributes very little
to the total mass of the universe: $\Omega_{lum} \approx 3.5\times
10^{-3} h_{100}^{-1}$.
\item{}Primordial nucleosynthesis gives for the {\it total} amount of
baryons $\Omega_{bar}^{(NS)} = 4\times 10^{-3} \eta_{10} h_{100}^{-2}$
with $\eta_{10} = 1-7$. At the moment there is no consensus
whether $\eta$ is relatively high (near the upper bound) or
at the lower end of the permitted interval.
In the first case we should expect plenty of invisible baryons in the
universe while in the second all baryons may be visible.
\item{}Measurements of rotational velocities in gravitationally
bound systems give, depending on scale, $\Omega = 0.1-0.3$.
\item{}From the large scale flows follows $\Omega_{lsf}=
[b(0.74\pm 0.13)]^{5/3}$ with the biasing parameter $b=O(1)$.
\item{}From the lower bound on the universe age $t_U > 12\,Gyr$ and
$h_{100} > 0.7$ follows $\Omega_{matter} <0.2$.
\item{}Measurements of X-ray emission from rich galactic clusters give
$\Omega_{clustered} \leq 0.15 h_{100}^{-1/2} /(1+0.55h_{100}^{3/2})$.
\item{}Inflationary cosmology predicts $\Omega_{tot} = 1$ with a very
good accuracy.
\end{enumerate}

It is rather difficult to say now what is the best choice in this
situation. The answer would be different depending on the personal
prejudices of the author. To my mind the odds are the best for the
following: $\Omega_{bar} \approx \Omega_{vis} \approx 0.5\%$;
$\Omega_{DM} =0.1-0.2$, and $\Omega_{vac} = 0.8-0.9$ with
$\Omega_{tot} = 1$.

\section{Possible Forms of Dark Matter}

The simplest and most economical is to assume that all dark matter
consists of the usual known particles: protons, neutrons, and
electrons, without unnecessary introducing quantities according to
Occam. Unfortunately this idea of a large amount of dark baryons
contradicts the nucleosynthesis bound and the only visible way to
overcome this bound is to conceal the majority of baryons in
primordial black holes. It can be done but with a rather exotic
model of baryogenesis which gives a very big baryon asymmetry inside
relatively small bubbles \cite{ds}. When the temperature drops
below the proton mass, very big density perturbations evolve which
result in almost 100\% collaps of the regions with high baryon
asymmetry. As is argued in ref.\cite{ds} (see also \cite{cos})
if baryonic black holes
have been formed sufficiently early the bound (\ref{omegab1}) may not
be applicable and $\Omega_B$ close to 1 is permitted.

There are many direct astronomical restrictions on the baryonic dark
matter in all possible forms which have been reviewed in
lectures \cite{js}. If the dark baryons are in the form of black holes
the following limits are to be fulfilled. Black holes with
$M>10^6 M_\odot$ would lead to formation of too heavy galactic
nuclei. Black holes with $M>10^4 M_\odot$ would disrupt globular
clusters. Black holes with $M<100 M_\odot$, which have been formed in the
process of heavy star evolution, are excluded because the star-progenitors
would eject too much metals into interstellar medium. However primordial
black holes are not subject to the last bound and for them
$M<100 M_\odot$ is permitted. Moreover, as we mentioned above,
even the nucleosynthesis bound for them may be avoided.

 Another possible candidate for dark baryons are brown dwarfs, light
star-like bodies consisting of hydrogen and helium with the mass
$M< 0.08 M_\odot$, so that the reaction of hydrogen fusion is not
ignited. Dark baryonic matter could also be in the form of the so called
"Jupiters", planetary-type objects with  $M< 10^{-3} M_\odot$. Compact
stellar remnants: white dwarfs and neutron stars ($M = (0.4-2) M_\odot$)
are also viable candidates. And at last one can expect some dark baryons
to be in the form of the cold diffuse gas clouds.

During the last 3 years several
groups of astronomers (EROS, MACHOS, OGLE)
are looking in our galaxy
for dark star- or planetary-size objects  with
the mass in the range
$ (10^{-8} - 10^3) M_\odot$ using the gravitational
microlensing effect \cite{bp}. These objects have got the name MACHOS
which is the abbreviation for Massive Astrophysical Halo Objects. When
such an object happened to be on the line-of-sight between the observer
and a star the brightness of the star may considerably increase due
to gravitational focusing of light. The luminosity curve should
have a symmetric form as a function of time and should not depend upon the
light frequency. Around 100 events of this type have been
observed to the present time. Unfortunately it is difficult
to judge if these
events are induced just by the ordinary stars or by the invisible celestial
bodies which are looked for. The experiments are now in progress and more
data are being accumulated. Moreover the new series of observations
are now done in real time when the change of star brightness is
immediately registered, while previously the
analysis of the accumulated data was performed.
So hopefully in the nearest future a more decisive conclusion can be
made. Still even if all the observed events are induced by MACHOS
is seems very plausible that they cannot account for
all dark matter in the universe. For the recent review see e.g.
ref.\cite{bp1}.

If we confine ourselves to the particles which are known to
exist, the next simplest possibility for dark matter are massive
neutrinos with $m=O(10\,eV)$. Though we cannot predict theoretically
the magnitude of the neutrino mass, there is no known principle that
forbids nonzero $m_\nu$ and so it is quite natural to expect them to
be massive. Unfortunately the theory of large scale structure formation
excluded massive neutrinos as dominant bearers of dark matter at least
in simple models with scale free spectrum of initial perturbations.
It is an interesting question whether there exists a spectrum
of initial perturbations
which permits to describe the universe structure with neutrinos
and baryons only.

If baryons and massive neutrinos cannot account for the observed
$\Omega_{DM}$ there are still plenty of possibilities but now we have
to turn to hypothetical not yet discovered particles or new objects
or fields. The most popular now candidate for the dark matter
particle is the lightest supersymmetric particle (LSP). In simple
supersymmetric models this particle is stable due to conservation
of R-parity, $R= (-1)^{B+L+2S}$ where $B$ and $L$ are respectively
baryonic and leptonic charge and $S$ is the particle spin. So defined
$R$-parity has the value $(+1)$ for all usual particles and $(-1)$ for
all (not yet discovered) superpartners.

In models with low energy
supersymmetry the SUSY breaking scale is phenomenologically acceptable
in the interval $\Lambda_{SUSY} = (0.1-1) \,TeV$. With this scale of
supersymmetry breaking the theory naturally predicts the relic
abundance of LSP's near $\Omega = 1$. Indeed the
number density of the primordial LSP's which survived to the present
epoch can be roughly estimated as \cite{dz}:
\beq{
n_{LSP} = {n_\gamma \over \sigma v m_{LSP} m_{Pl} }
}\label{nlsp}
\eeq
where $\sigma$ is the cross-section of annihilation of LSP's,
$v$ is their velocity, and $n_\gamma = 400/cm^3$ is the number density
of photons in cosmic microwave background. With $\sigma v =
\alpha^2 /m_{LSP}^2$ where $\alpha \approx 10^{-2}$ and
$m_{LSP} \approx \Lambda_{SUSY}$ we get
\beq{
\Omega_{LSP} \approx 0.04 h^{-2}_{100} (m_{LSP} /TeV)^2
}\label{omegalsp}
\eeq
More details about and a review on supersymmetric dark matter
can be found e.g. in paper \cite{kane}.

However the conservation of R-parity is not obligatory. No theoretical
principle like e.g. gauge invariance demands that. So it is quite
natural to expect that it is indeed broken. Moreover if we believe
the principle, which has been established by the development of the
elementary particle physics during the last 30 years,
that everything that can be broken should be broken,
R-parity is surely broken. If this is the case the LSP should
be unstable. Still depending on the mechanism and the strength of
breaking the life-time of LSP's can be very long like e.g. the
life-time of proton which is stable in all known experiments and
observations despite an almost certain baryonic charge nonconservation.
If so, the long-lived LSP can well be the dark matter particle. If
however their life-time is cosmologically short, they could not
contribute to dark matter and other candidates are wanted.

In models with broken R-parity an interesting candidate for dark
matter particles can be massive majorons \cite{bv,dpv} i.e
pseudogoldstone bosons appearing due to spontaneous breaking of
leptonic charge conservation. The majorons may have mass in KeV range
and the number density corresponding to $\Omega = O(1)$. The universe
structure formation in such model has been considered in ref.\cite{dpv}
and reasonably well agrees with the observed picture
(see section 12).

There could exist some other stable (or cosmologically long-lived)
particles that may be bearers of dark matter. They could be produced
thermally, like LSP's or (massive) neutrinos  in hot equilibrium
primeval plasma. Depending on the strength of their interactions and
the magnitude of the mass they may
be relativistic when they are decoupled from the cosmic plasma
(e.g. neutrinos) or nonrelativistic (e.g. LSP's). Another possible
mechanism of production may be related to a phase transition and
in this case even very light particles would be produced in
nonrelativistic state. An example for that is the well known axion
which is also considered as a serious candidate for dark matter
particle. The effective potential for axion field has the form
of Mexican hat after the underlying $U(1)$-symmetry is broken
at high temperature. Later on at QCD energy scale the bottom
of the Mexican hat potential became tilted and because of that
axions acquire a nonzero mass. At that moment coherent
oscillations near the potential minimum begin. This can
be understood as a production of Bose condensate of axions
(at rest). Though the mass of axion is extremely small,
$m_a = O(10^{-5}\, eV)$, the axions, thus produced, remain
nonrelativistic because of negligibly weak interactions
with the plasma.

Some other, maybe more exotic, and because of that more interesting,
candidates for dark matter are considered in the literature
like e.g. topological or nontopological solitons with astronomically
large size. Topological defects (or topological solitons)
might be formed in the course of phase transitions
accompanying spontaneous symmetry breaking when the universe
expanded and cooled down. Depending upon the symmetry group
different kinds of objects can be created. If the corresponding
symmetry group is $U(1)$ then vortex lines (or cosmic strings)
would be formed which
could both make contribution into dark matter and
create density inhomogeneities \cite{ybz,av}. If a discrete
symmetry is broken domain walls separating different vacua
are created. Domain walls with a natural (whatever it means)
scale are cosmologically forbidden since they would create
too large angular fluctuations of temperature of cosmic microwave
background \cite{koz} but if the symmetry breaking scale is
extremely low, they are allowed and can even
serve as seeds for structure formation \cite{dns}. Spontaneous
breaking of $SU(2)$ or $O(3)$ gives rise to production of
monopoles. These three types of objects exhaust all possible
topologically stable solitons in three dimensions. There could also
be unstable but long-lived ones which might be cosmologically
interesting. For the review see e.g. refs. \cite{av2,ad5}.

I think that a very interesting possibility for dark matter is
a classical field which possibly compensates nonzero vacuum
energy. The theory of such field has not yet been seriously
considered (even at a toy model level) but the cosmological
constant problem makes the existence of such a field very
probable.

At the present level of our knowledge we cannot say what is the
theoretically preferable candidate for dark mater particles. We only
can make a judgement on the basis of the universe structure formation
comparing the picture predicted by a model with a particular kind of
dark matter particles and astronomical observations.

\section{Structure Formation. Basic Assumptions.}

The universe is believed to be homogeneous on the average at very
large scales, $l>l_1 \approx 100\, Mpc$. This hypothesis is based
on the observed isotropy of the cosmic microwave background and on
the absence of the observed structures at $l>l_1$. However the
present-day accuracy of measurements is not good enough to exclude
structures at larger scales so the last statement may be questioned.
At smaller scales the universe is very inhomogeneous: there are
galaxies, their clusters and superclusters, large size voids, and
possibly even a periodicity (or at least structures) with
the characteristic wave length about 100 Mpc \cite{period}.

The task of the theory of large scale structure formation is to
give a qualitative description of the distribution of matter in the
universe. This is done
with the following four essential assumptions of a different level of
creditability:

1. It is assumed that the structure of the universe has been formed
as a result of evolution of initially small fluctuations under the
action of gravitational forces. This is the only solid assumption of the
theory with which practically everybody agrees. Still even this
sometimes is questioned and speculations about possible nongravitational
processes or initially large fluctuations are put forward.

2. The next very important assumption concerns the nature and the
spectrum of initial perturbations. For many years the origin
of primordial fluctuations at very large scales remained mysterious.
A natural idea that density inhomogeneities were generated by
quantum or thermal fluctuations could not be realized in frameworks
of the Friedman cosmology with the power law expansion regime,
$a(t) \sim t^{\alpha}$, because the corresponding wave length
were negligibly small in comparison with the necessary astronomical
scales. This difficulty was resolved in inflationary
scenario \cite{gp} where
the wave lengths of perturbations were exponentially inflated,
$l_0 \rightarrow l_0 \exp (H\tau)$, and what's more their amplitudes
were amplified in the course of expansion. In a sense the plan
was overfulfilled because the density perturbations generated in
this way would be too high unless an unnaturally weak interaction
of the inflaton field is assumed.
It can be shown that for the inflaton with the self-interaction
potential $U(\phi) = \lambda \phi^4$ the magnitude of density
perturbations with wave length $l$ at the moment of horizon crossing
is roughly given by the expression:
\beq{
\delta = {\delta \rho \over \rho } \approx 100\sqrt \lambda
   [1+0.01\ln (l/l_{gal})]^{3/2}
}\label{deltainf}
\eeq
For the details see either the original papers \cite{gp}
or the book \cite{linde}. Note that to create the necessary
magnitude of fluctuations, $\delta < 10^{-4}$, bounded by
the data on the angular fluctuation of the background radiation
temperature, one needs $\lambda < 10^{-12}$.

A competing source of density perturbations may be astronomically
large topological defects mentioned in section 9. The simplest and
the most attractive possibility to my mind are cosmic strings.
Unfortunately the accepted $SU(3)\times SU(2) \times U(1)$-theory
of particle interactions does not lead to cosmic strings
(neither to any other mentioned above stable topological defects) and
at the present level of our knowledge we cannot say if higher energy
modifications of the theory definitely request existence of such
objects.

If we neglect the log-dependence on the scale,
the spectrum of fluctuations given by eq.(\ref{deltainf}), coincides with
that proposed by Harrison \cite{har} and Zeldovich \cite{ybz2}
some time before inflationary cosmology was discovered.
This is the so-called
flat or scale-free spectrum of fluctuations. It is the simplest possible
primordial spectrum which does not introduce any particular scale to
the theory. With  an advent of inflation this kind of spectrum got
theoretical justification and now is commonly used as the basic spectrum
in calculations of structure formation. Sometimes as a simple
generalization an  arbitrary power law spectrum is considered:
\beq{
\delta^2 = \left({\delta \rho \over \rho }\right)^2
\sim \int {dk\over k^n}
}\label{deltan}
\eeq
The Harrison-Zeldovich spectrum corresponds to $n=1$. The spectrum with
$n\neq 1$ may appear in some inflationary models \cite{kf} but with $n$
not much different from unity. If we permit in principle an existence
of primordial spectrum with an arbitrary $n$, introducing in this
way a new scale to the theory, a combination of several
terms with different $n$ as well as a more complicated function
are also permitted but without a guiding principle for choosing a
particular form of the spectrum, the theory would completely loose
its predictive power.

3. Because of stochastic nature of the fluctuations one should
consider average values and this explains in particular the
consideration of $\delta^2$ instead of $\delta$ in eq.(\ref{deltan}).
Stochastic properties of $\delta (x)$ are fixed by the following
hypothesis. It is assumed that the fluctuations are gaussian
with delta-correlated Fourier modes. In other words if the relative
amplitude of the fluctuations is Fourier expanded:
\beq{
\delta (x) \equiv {\delta \rho (x) \over \rho } =
\int {d^3k \over (2\pi)^3} \bar\delta(\vec k) e^{i\vec k \vec x}
}\label{deltax}
\eeq
the averaged product of the mode amplitudes satisfies the condition
\beq{
\langle \bar\delta(\vec k) \bar\delta(\vec q) \rangle =
(2\pi)^3 f(\vec k)^2 \delta (\vec k - \vec q )
}\label{deltadelta}
\eeq
(hopefully the delta-function here is not confused with the relative
magnitude of density fluctuations $\delta(\vec x)$ and its Fourier
transform $\bar \delta (\vec k)$).

It is also assumed that the spectrum is isotropic so that the function
$f$ is a function of the absolute value of $k$ only,
$f(\vec k ) = f(k)$. (Note in passing that this assumption is not
obligatory but we do not know any reasonable physical mechanism for the
anisotropy.)

Using these equations we can easily find
\beq{
\langle \delta^2 (x) \rangle = \int d^3 k f^2(k)
}\label{delta2}
\eeq
With $f^2 = const/ k^{n+2}$ it coincides with eq.(\ref{deltan}).

4. Except for the three hypothesis discussed above, the structure formation
evidently and heavily depends upon the properties of matter which presents
building blocks for the structures. In contrast to the previous three
assumptions there is no consensus with respect to the universe
chemical content and a plethora of different possibilities is
proposed. From the point of view of structure formation all possible
forms of matter particles are classified as cold, hot, and,
for the intermediate case, warm. Cold matter or CDM (here "D" stands
for "dark" because it is assumed that the universe is dominated by dark
matter) is formed by particles
which were nonrelativistic at a rather early stage
when the galactic size crossed the horizon.
Hot matter (or HDM) is formed by particles which were relativistic at
that moment, and correspondingly warm particles were semirelativistic.
This property of dark matter particle determines the characteristic size
of objects formed.

At the present day there are several models which more or less
successfully describe structure formation under the first three
assumptions and with different hypotheses about the matter content of
the universe: CDM plus vacuum energy \cite{vacstruc}, mixed hot plus
cold dark matter \cite{mixed}, unstable particles decaying into
relativistic species plus CDM \cite{unstable}, HDM plus astronomically
large seeds like topological defects. These models request comparable
contributions of different forms of matter which are not physically
related at least at the present level of our understanding and by
this reason do not look natural. On the other hand we already have
an example of such unnaturalness realised in the universe,
namely the energy density of baryons is relatively close to that
of dark matter (compare $\Omega_{bar}$ with $\Omega_{DM}$) though they
may be different by several (many!) orders of magnitude. At the moment
it is not clear if this rough coincidence is an indication of an
unknown deep physics, or there is something wrong in our understanding
of dark matter, or we have to invoke the anthropic principle to
explain that. Models with a single dominant form of matter demand
a deviation from the flat spectrum of primordial density fluctuations.

It is usually assumed that the dark matter
is collisionless (or in other words its selfinteraction as well as
interactions with other particles may be neglected)
but this is not necessarily so and the models with self-interacting dark
matter are also considered \cite{hall,mem,dpv}.

We will briefly discuss the relevant properties of models of structure
formations with different forms of dark matter particles or fields
below.

\section{Structure Formation. Basics of the Theory.}

The evolution of small density perturbations is described by  the usual
hydrodynamic equations of selfgravitating perfect fluid with some
complications related to the expanding cosmological background and
effects of general relativity. The dynamics of liquid element is
governed by the forces of gravity and pressure. If the pressure
dominates then the perturbations evolve as sound waves. For the
collisionless liquid like e.g. that of neutrinos the perturbations are
erased by free-streaming which formally very much resembles the
effect of pressure. In the case
when gravity dominates, the element of liquid
would be indefinitely compressing if the equation of state remains
unchanged and correspondingly no new pressure forces come into play.
In this case the density contrast would be rising and this results
in formation of structures in the universe. Evidently with an
increasing size or mass of a system
gravity would ultimately dominate pressure while for small
systems pressure is normally the dominant force. The boundary value of
the mass and size are called respectively Jeans mass, $M_J$, and
Jeans wave length, $\lambda_J$. They are named so after Jeans
who was the first to study the phenomenon of gravitational instability
at the beginning of this century.

The basic equations which describe evolution of density
perturbations in the nonrelativistic perfect fluid with mass density
$\rho$ and pressure density $p$ are the continuity equation:
\beq{
\partial_t \rho + \vec \partial (\rho \vec v) = 0,
}\label{contin}
\eeq
the Newtonian equation of motion ($m\ddot {\vec x} = \vec F$):
\beq{
\partial_t \vec v + (\vec v \vec\partial)\vec v +
   \vec\partial p /\rho + \vec\partial\phi = 0,
}\label{vdot}
\eeq
and the Newtonian equation for the gravitational potential $\phi$:
\beq{
\Delta \phi = 4\pi \rho G_N
}\label{deltaphi1}
\eeq
Perturbative expansion of these equations over homogeneous
background $\rho = \rho_0  + \rho_1 (t, \vec x)$,
$\vec v = \vec v_1 (\vec x ,t)$, etc, permits to derive
the second order linear differential equation describing evolution
of density perturbations:
\beq{
\ddot \delta - v^2_s \Delta \delta - 4\pi \rho_0 G_N \delta = 0
}\label{ddotdelta1}
\eeq
where $v_s$ is the velocity of sound.

Making Fourier transform of this equation
one sees that short wave modes are oscillating
while the long wave ones are exponentially rising,
$\delta \sim \exp \left(\sqrt{4\pi G_N \rho_0 - v^2_s k^2}\,t\right)$.
The boundary
value of the wave vector $k_J^2 = 4\pi G_N \rho_0 /v^2_s$ is called
the Jeans wave vector and the inverse quantity $\lambda_J = 2\pi /k_J$
is the Jeans wave length. The Jeans mass is the mass contained
inside the volume bounded by $\lambda_J$ and is equal to
$M_J=4\pi (\lambda_J /2)^3 \rho_0 /3 =
\pi^{5/2} v_s^3 / 6G_N^{3/2} \rho_0^{1/2}$.
In realistic cosmological situation these expressions remain true with
the substitution the cosmological
energy density $\rho_0 (t) \sim 1/t^2$
instead of $\rho_0 = const$ . Because of that the character
of instability becomes different, it is no more exponential but of
a power law as we have seen in sec.5.

To derive the equations of motion in Newtonian approximation with
the account of the universe expansion one has to change all the
derivatives to the covariant ones in the Robertson-Walker metric,
to substitute as zero order approximation the cosmological time
dependent energy density $\rho_0 (t)$ (as we have already mentioned)
and the nonvanishing zero order velocity, $\vec v_0 = H \vec r$.
Otherwise the procedure is exactly the same as
above. The resulting equation for the Fourier transformed
amplitude of density perturbations $\delta_k$ is similar to eq.
(\ref{ddotdelta1}) with an extra "friction" term proportional to
$H=\dot a /a $:
\beq{
\ddot \delta_k + 2 (\dot a/a) \dot \delta_k + (v^2_s k^2 /a^2 -
4\pi G_N \Omega \rho_0) \delta_k = 0
}\label{ddotdeltak}
\eeq
The limit of small $k/a$ of this equation gives eq.(\ref{ddotdelta}).
More details about the material presented above and below
can be found in the books \cite{peeb,kt} or in the
lectures \cite{efst,dhl}.

It is essential for structure formation that perturbations in
nonrelativistic matter do not grow in the universe dominated by a
{\it uniformly} distributed relativistic matter. In this case
the term which drives the instabily in eq.(\ref{ddotdeltak}) is
changed to $4\pi \rho_0 G_N \epsilon $ where $\epsilon =
\rho_m/\rho_0 $, $\rho_m$ is the energy density of nonrelativistic
matter, and $\rho_0$ is the dominant cosmological energy density of
relativistic matter. Since by assumption $\epsilon \ll 1$, the driving
force becomes negligible and rising modes are not generated. If the
relativistic matter is itself inhomogeneous then there are unstable
modes but usually inhomogeneities in relativistic matter are quickly
erased by free streaming or diffusion. Correspondingly cosmological
inhomogeneities started to grow only at matter dominated stage.

General relativistic treatment of the problem is more
complicated. One has to solve the Einstein equations for the metric
with corrections from inhomogeneous fluctuations. There is an
ambiguity related to the freedom in choice of different coordinate
systems (the so called gauge freedom), so that one has to
distinguish between the gauge artifacts and real physical effects.
This analysis was first done by
Lifshits in 1946\cite{eml}. The gauge independent
formalism was developed by Bardeen \cite{bar} relatively recently.
In many cases however it is sufficient to use the
Newtonian approximation for qualitative (and often quantitative)
understanding of the phenomena.

While the analysis of the evolution of small perturbations is relatively
simple the nonperturbative regime which started when $\delta = O(1)$
is much more complicated. No reliable analytical or semianalitycal
methods have yet been developed and the calculations are made by
numerical simulations with a large number of noninteracting (except
for gravity) particles. By necessity many essential properties of the
system are neglected but nevertheless one may hope that this approach
gives a reasonable description of the structure formation at the
stage of large perturbations.

\section{Evolution of Perturbations with Different forms of Matter.}

Let us first consider density perturbations in the cosmological gas
of light neutrinos with mass in 10 eV range. At an early stage when
neutrinos are relativistic density contrasts are smoothed down by
their free motion. The perturbations could start rising only when
neutrino momentum is redshifted down to nonrelativistic values. The
characteristic size of the fluctuations which survived the erasure by
free streaming, is given by the path which neutrinos could travel
till they became nonrelativistic, $l_m\approx 2t_m$. Here $t_m$ is the
universe age at that moment. It is assumed that neutrinos moved with
the speed of light and the coefficient 2 is connected with the
universe expansion (in the static universe the path would be just
$l=t$). We find $t_m$ using equations (\ref{rhoc1},\ref{rhot}):
$t_m = 4\times 10^{20} \,cm (10\,eV /T_m)^2$. We choose $T_m =
m_\nu /3$ because the average neutrino energy in thermal equilibrium
state is $\langle E \rangle \approx 3T$. To calculate the magnitude
of $l_m$ at the present day we have to multiply it by the red shift
$z_m + 1 = T_m / T^{0}_\nu$ where $T^{0}_\nu = 1.93 \,K$ is the
present day temperature of the cosmic massless neutrinos. Thus we get
\beq{
l_m^{(0)} = 16 \,Mpc (10\,eV /T_m) = 48 \,Mpc (10\,eV / m_\nu)
}\label{l0m}
\eeq
The mass of matter inside the radius $l_m/2$ is $M_m \approx
3\times 10^{15} M_\odot (10\,eV /m_\nu)^2$. It is 3-4 orders of
magnitude bigger than galactic mass. Note that $l_m^{(0)}$ and
$M_m$ coincide correspondingly with $\lambda_J$ and $M_J$ given
above if we formally take the speed of sound equal to the speed of
light. This is indeed true for relativistic gas. Eq.(\ref{l0m})
gives the mass of the smallest size objects which could be
initially formed in neutrino dominated universe. Smaller
objects is difficult to create and this was the reason why a beautiful
idea of neutrino universe was abandoned.
Particles with a larger mass like e.g. $m= 300 \,eV$ would create
structures with characteristic size close to that of galaxies. As we
mentioned above these are particles of warm dark matter.
There is a renewed interest to these particles now.
Unfortunately no particles with such mass are observed experimentally
but there are some theoretical models (see e.g. refs.\cite{bv,dpv})
where such particles may exist and contribute to dark matter.

During several last years, before the COBE measurements \cite{cobe}
of the large angle anisotropy of the cosmic microwave background
radiation, the model with cold dark matter was definitely
the most popular (for the review see e.g.
refs. \cite{mixed,cdm}).
Based on a very simple assumption of scale-free spectrum of
primordial fluctuations (Harrison-Zeldovich  spectrum), the model
reasonably well described the galaxy distribution at the scale of tens
megaparsec with the only free parameter, the overall normalization
of the spectrum. However the COBE measurements have
fixed the normalization at large scale end of the spectrum and with
this normalization the prediction for smaller scales
became a factor two
above the observations. An evident possible cure is to change the
shape of the spectrum of primordial fluctuations. It seems to be
a very interesting possibility but since this idea is not
supported by the inflationary scenario, the major line of investigation
is the consideration of different forms of dark matter
with the same flat spectrum of density perturbations or maybe an
introduction of the cosmological constant.
The basic idea of all these models was to
suppress the power of the evolved spectrum at smaller scales relative
to that at larger (COBE) scales.

One possibility is to shift the epoch of matter dominance (MD)
to a later stage in a simple cold dark matter model.
Since the characteristic scale at which perturbations started to rise
in this model is determined by the horizon size at the onset of the
MD stage, shifting it to a later moment gives less time for rising of
the fluctuations and correspondingly less power at galactic
and cluster scales. This goal can be achieved if one assumed
that universe is open so that
$h_{100}^2 \Omega \approx 0.2 $. This model
is disfavoured by inflationary scenarios which (at least in simple
versions) predicts $\Omega = 1$. One can recover this prediction
of inflation in the universe with low matter density if the
cosmological constant $\Lambda$ (or in other words, vacuum energy) is
nonzero. As we have mentioned in sec. 2,
the recent data indicating a rather high value of
$h_{100} \approx 0.8$ support the idea of nonzero $\Lambda$
with the fraction of the vacuum energy $\Omega_{vac} = 0.8$.
The models with nonzero
$\Lambda$ give a satisfactory description of the observed
structure \cite{vacstruc} with flat spectrum and COBE normalization.

A mixed (hot+cold) dark matter scenario can
also do the necessary job of diminishing the power at small
scales because
(initially flat) perturbations in hot dark matter are erased at
scales smaller than $\sim 10^{14} M_{\odot}$ by free streaming if the
dark matter is collisionless as is the case of neutrinos. A good
description of the structure requests 70\% of CDM and
30\% of HDM \cite{mixed}. It
gives even better description if there are two equal mass neutrino
species each with the mass 2.5 eV \cite{prim} as is suggested by
the recent indications of neutrino oscillations by the Los Alamos group.

Recently there appeared a renewed interest to the idea of
structure formation with unstable particles
\cite{unstable}. It is assumed in these
models that there exists a massive long-lived particle, usually
tau-neutrino with the mass in MeV range which decayed into massless
species at the epoch
when the mass density of the parent particles dominated the
energy density of the universe. Correspondingly the present-day
energy density of relativistic particles would be bigger than in the
standard scenario and the onset of MD stage would take place later.

The common shortcoming of these models is that they all demand a
certain amount of fine-tuning. Generally one would not expect that
the contribution from hot and cold dark matter into the universe
mass are about the same, they may differ by many
orders of magnitude.
One would also suspect that the vacuum energy which remains
constant in the course of the universe expansion is by no means related
to the critical energy now (which goes down with time as
$m_{Pl}^2 /t^2$)
\footnote{Adjustment mechanism \cite{ad3,ad4} (for more references
see reviews \cite{sw2,ad2} and the recent paper \cite{yf}) however
generically gives the noncompensated amount of vacuum energy of
exactly necessary order $m_{Pl}^2 / t^2$ and so the models with
vacuum energy may be more "equal than others". }.
The models with unstable particles mentioned above are also based on
the assumption of two independent components: the massive unstable
particles themselves and unrelated cold dark matter. This is
definitely unnatural and this shortcoming stimulated search for
other models. Recently in ref. \cite{wssd} a return to to the
model with a single cold dark matter component universe (of course
except for baryons) was advocated. For successful description
of the observed structure the authors need the power index of the
spectrum $n=0.8-0.9$, a low value of the Hubble constant,
$h_{100} = 0.45-0.5$ (to be compatible with large $\Omega$), and
a large contribution of tensor perturbations (gravitational waves)
into quadrupole fluctuation of the background radiation temperature
measured by COBE, $C_T /C_S = 0.7$.

Another attempt to overcome unnaturalness of multicomponent dark
matter model has been done in paper \cite{dpv}.
In this paper a model is considered in which unstable particles and
the particles of cold (or possibly warm) dark matter are closely
connected. In fact the decay of the former produces particles of the
present-day dark matter. A necessary background model of this kind
in particle physics was proposed some time ago \cite{jv1,CMP} as
an attempt to find a phenomenologically acceptable description of
R-parity breaking. The underlying mechanism is the
spontaneous breaking of leptonic charge conservation at
electroweak scale. The model contains a Majorana type
tau-neutrino with the mass around MeV which decays into massive but
light Majoron $J$ with mass in KeV region. Life-time with respect to
this decay, as estimated in refs. \cite{jv1}, could be of order days or
years depending on the values of parameters, i.e. just in the interesting
for us interval. It is worth noting that there is no
stable SUSY particle in this model
so the dark matter cannot be associated with it but the model itself
produce a candidate for dark matter, namely a massive Majoron.
The model possesses some features like
sufficiently large diagonal coupling of $\nu_\tau$ to majorons or
selfinteraction of majorons which  rather naturally permit
to resolve appearing cosmological problems in particular the
problem of extra massless particle species during primordial
nucleosynthesis. The cosmological properties
of this model are rather unusual and are interesting by themselves.
Dark matter particles (majorons) in this model are strongly
self-interacting and thus the structure formation in this model
is different from the traditional one with collisionless dark matter
particles \cite{hall,mem,lss,dpv}. In particular the shape of galactic halo
would depend on the dark matter selfinteraction. The
recent data \cite{shape} might give an indication that collisionless
dark matter gives a poor description of the shape of halo in
dwarf galaxies.

\section{Conclusion.}

It is difficult to write a conclusion on something which is so
evasive as dark matter. Though it is practically certain that dark
matter exists, it is unknown what it is and even if there are one or
more forms of dark matter. So at the present level of our knowledge
we can only make bets which are very little constrained by the poor
available data.

One of the most important questions is whether $\Omega$ is indeed equal
to 1 as predicted by inflation or it is considerably smaller. In the
last case the universe age problem is not so severe with a large
Hubble constant but still if $h_{100} > 0.75$ and $t_U > 14\,Gyr$
even $\Omega =0$ does not help much. It is very important to know
$H$ more accurately since its value may give a weighty argument in
favor of nonzero cosmological constant $\Lambda$. With nonzero
cosmological constant $\Omega_{matter}$ could be relatively small
(while $\Omega_{tot}= \Omega_{matter}+\Omega_{vac}=1$) and the universe
age problem does not exist. Structure formation in models with
$\Lambda \neq 0$ looks reasonable too and the unnaturally close values
of $\rho_c$ and $\rho_{vac}$ may be understood with an (unknown)
adjustment mechanism. So one of the most important problems is to find
a workable model for the latter or in more general terms to solve
the cosmological constant mystery. If it is indeed solved by an
adjustment mechanism one should expect that $\delta\rho_{vac}
\approx \rho_c$ during all the history of the universe. Hence the
cosmology with adjustment of vacuum energy may give somewhat
different predictions for light element abundance, structure formation,
etc. Probably the baryonic crisis is also resolved by a nonzero
cosmological constant.

Speaking about more traditional forms of dark matter we have to note
that nonzero $\rho_{vac}$ is not sufficient to solve all problems
with dark matter and some amount of "normal" dark matter is necessary.
What is it is a big question. It may be LSP's or axions, or massive
neutrinos. Though the latter are disfavored by the theory of structure
formation, it may be premature to exclude them completely
as a dominant dark matter (together with vacuum energy). If R-parity is
not conserved LSP's are probably excluded but a massive majoron in this
case becomes a very interesting candidate. Accelerator experiments in
search of supersymmetry are very important for resolution of this
problem.

Invisible baryons may also exist and to this end a clarification of
situation with primordial nucleosynthesis is very desirable
especially a resolution of the ambiguity with the observation of
primordial deuterium. In particular it may settle down the problem
with the number of massless particle species present at nucleosynthesis
and exclude or confirm a hypothesis of MeV-tau-neutrino.

In a whole we can expect a really exciting development in the nearest
future.

{\bf Acknowledgements}

This work was supported by DGICYT under grants
PB92-0084 and SAB94-0089 (A. D.).

\end{document}